\documentclass[9pt,twocolumn,twoside]{opticajnl}
\journal{opticajournal} 

\setboolean{shortarticle}{true}

\usepackage{lineno}
\usepackage{caption}
\usepackage{subfig}
\usepackage{array,calc}
\usepackage{float}
\usepackage{graphicx}
\usepackage{subcaption}

\title{Purcell Enhancement of Spontaneous Emission of a Quantum Emitter on a Waveguide}

\author[1]{SUSHMA GALI}
\author[2]{KOMAL SHARMA}
\author[2]{JAYDEEP KUMAR BASU}
\author[1,*]{SHANKAR KUMAR SELVARAJA}

\affil[1]{Centre for Nanoscience and Engineering, Indian Institute of Science, Bengaluru,560012, India}
\affil[2]{Department of Physics, Indian Institute of Science, Bengaluru,560012, India}

\affil[*]{Corresponding author: shankarks@iisc.ac.in}

\begin{abstract}
We investigate the effect of a waveguide on an emitter's spontaneous emission in its vicinity. The impact of various possible orientations of an emitter with respect to the waveguide surface is studied through simulations and compared with experimental demonstration. Quantum emitters are dip-coated on waveguides, and Purcell enhancement and a decrease in the lifetime of the emitter are observed. This study serves as a proof of concept for the Purcell effect offered by waveguides and helps in effectively estimating the efficiency of waveguide-based evanescent sensors and quantum photonic applications
\end{abstract}

\setboolean{displaycopyright}{false} 

\begin{document}

\maketitle

\section{Introduction}
Integrating quantum dots with waveguides represents an interesting convergence of quantum optics and photonics, offering exciting opportunities for manipulating and controlling light at the nanoscale. Due to their confined size, quantum dots exhibit unique quantum mechanical properties, including discrete energy levels. When coupled with waveguides, which are structures designed to guide and manipulate light flow, these quantum dots become powerful elements for advanced optical and photonic applications. The interaction between quantum dots and waveguides allows for precise control of emission, absorption, and transmission of photons. Integration of quantum dots with waveguides is instrumental in the development of efficient quantum light emitters that are essential for on-chip quantum information processing [1],[2],[8]. In this regard, we study the effects of waveguides on the emission properties of quantum dots sitting on them. A quantum dot can be treated as a dipole emitter, and the spontaneous decay rate of a dipole oscillating at frequency $\omega$ in a homogeneous isotropic medium of refractive index $n_i$ can be given by [12]
\begin{equation}
    \gamma_0 = \frac{n_i \omega^3 |\mu|^2}{3 \pi \hbar\epsilon_0 c^3}
    \label{eq:1}
\end{equation}
Where $\mu$ is the dipole moment of the emitter, the spontaneous emission rate (also the excitation lifetime) of an emitter depends on the structures and refractive index of the surrounding medium. In an inhomogeneous medium, there could be an enhancement in the spontaneous emission of the emitter; this effect was first reported by Purcell and is named after him as Purcell enhancement [3-6]. To estimate the Purcell factor in a waveguide-emitter system, one needs to start with Fermi's golden rule [7-9], which gives a relation between the decay rate of an emitter in terms of oscillation frequency $\omega$ and position $r$.
\begin{equation}
    \gamma_c = \frac{2\pi}{\hbar^2}|\tau(r,\omega)|^2N(r,\omega)
    \label{eq:2}
\end{equation}
Where $\tau(r,\omega)$ is the coupling strength between the dipole and electromagnetic field, $N(r,\omega)$ is the density of states corresponding to $h\omega$. The ratio of Eq. \ref{eq:1} and Eq. \ref{eq:2} gives the enhancement in the spontaneous emitter emission in the presence of a waveguide and can be written as shown in  Eq. \ref{eq:3}. Detailed supporting data that help arrive at  Eq. \ref{eq:3} can be found in [8-11].
\begin{equation}
    \frac{\gamma_c}{\gamma_0} = \frac{3c\lambda_0^2}{2n_i^3}\frac{1}{LA_eff} N(r,\omega)
    \label{eq:3}
\end{equation}
Where $A_eff$ is the effective mode area of the waveguide and $L$ is the quantization length. The density of states $N(r,\omega)$  And quantization length can be related to group velocity $v_g$ of mode as
\begin{equation}
    N(r,\omega) = \frac{L}{\pi v_g}
    \label{eq:4}
\end{equation}
Now, the Purcell factor P can be written as
\begin{equation}
    P = \frac{\gamma_c}{\gamma_0} = \frac{3}{2\pi}\frac{C}{v_g n_i^3}\frac{\lambda_0^2}{A_eff}
    \label{eq:5}
\end{equation}
$P$ should be greater than one for an emitter to show enhanced spontaneous emission. In the case of a waveguide-emitter environment, the waveguide structure should have a propagating mode with a modal area $A_eff$ and group velocity $v_g$ to enhance emission (Eq. \ref{eq:5}). The Purcell factor for a given waveguide dimensions can be calculated from the FDTD simulation of effective modal area and group velocity.

In this paper, we report n

investigated the effect  variations in the Purcell factor of a quantum dot on a waveguide concerning factors such as the distance from the waveguide, waveguide width, and emitter orientation. The Purcell factors obtained from simulations were further compared with experimental demonstrations for comprehensive analysis.

\section{simulations}
An emitter approximated as a dipole molecule, can be randomly oriented when dip-coated on a waveguide surface. But to understand the effect of dipole orientation on Purcell enhancement, three possible dipole orientations, normal, parallel, and tangential, are considered, as shown in figure1. A 220 nm thick Silicon Nitride (SiN) waveguide on two-micron buried oxide is considered for simulation study as SiN waveguides are promising for on-chip spectroscopic applications [1],[2]. The Purcell factor described in equation (5) is extracted by placing a dipole source on the waveguide and at the side wall of the waveguide. Figures 2(a) and 3(a) show the dipole placement environment for simulation. Figures 2(b) and 3(b) show how the Purcell factor varies with the orientation and position of the dipole emitter. The simulation results shown in Figure 2 and Figure 3 essentially tell that if the dipole emitter axis is in the plane with the cross-section of the waveguide, more Purcell enhancement can occur.

\begin{figure}[ht]
\centering
\includegraphics[width=\linewidth]{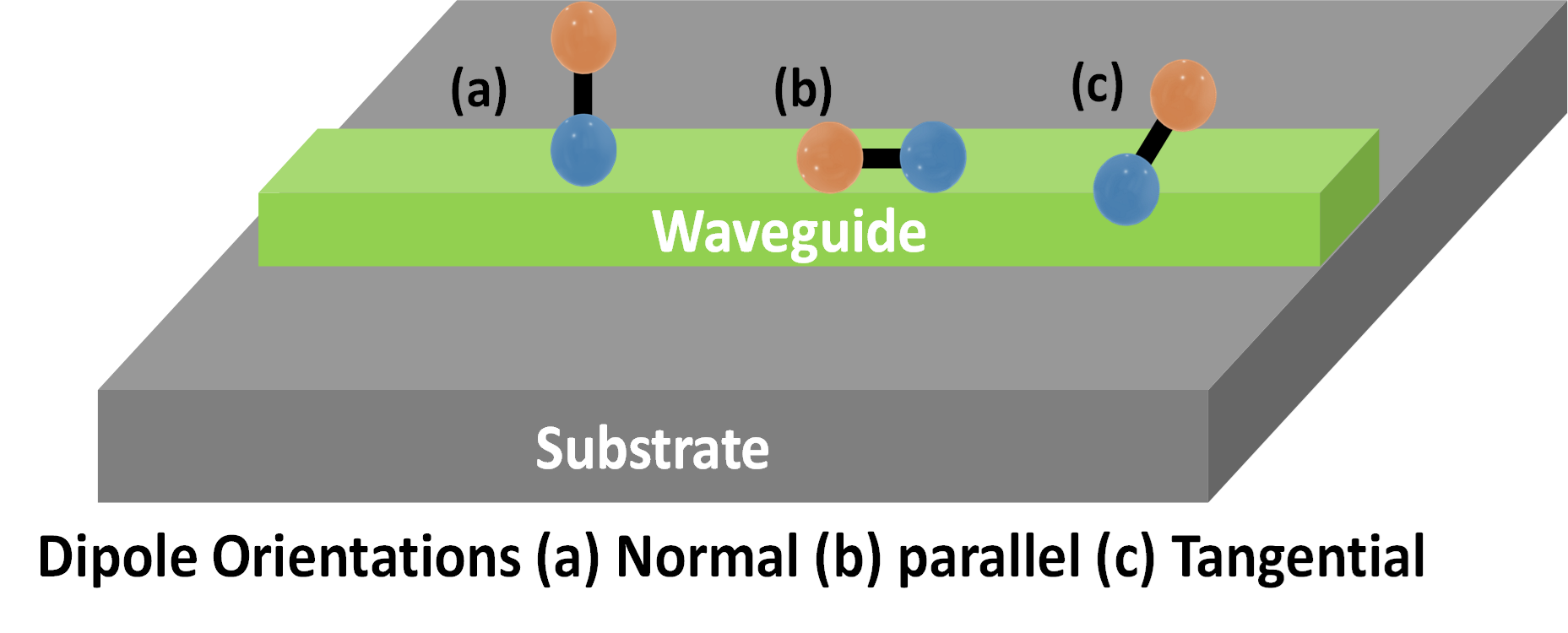}
\caption{Various orientations of a dipole emitter on a waveguide}
\label{fig:1}
\end{figure}









\begin{figure}[ht]
\centering
\includegraphics[width=\linewidth]{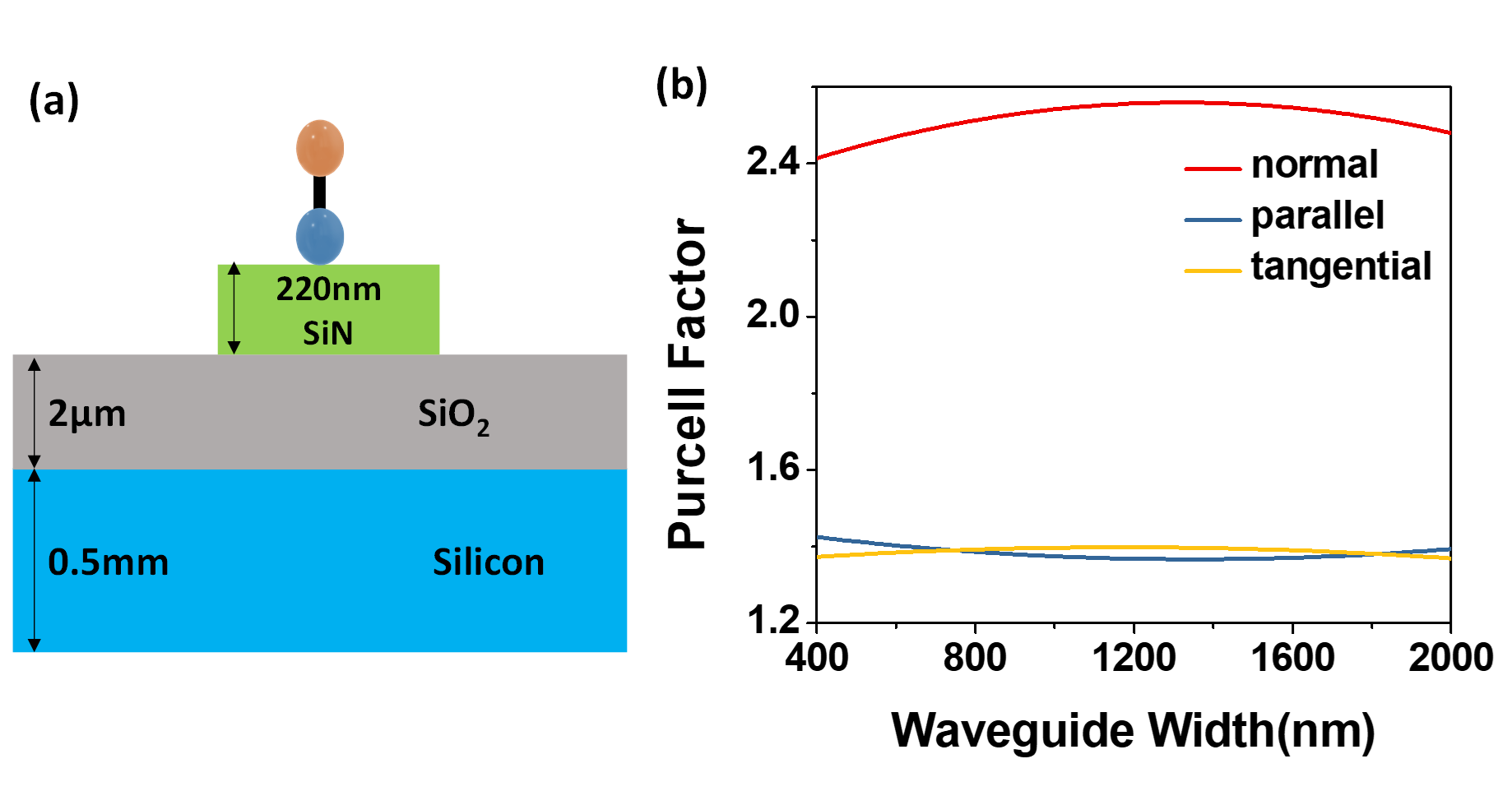}
\caption{ (a) Cross section of SiN waveguide with dipole emitter on its surface
 and (b) Purcell factors of various dipole orientations with change in waveguide width
}
\label{fig:2}
\end{figure}

\begin{figure}[ht]
\centering
\includegraphics[width=\linewidth]{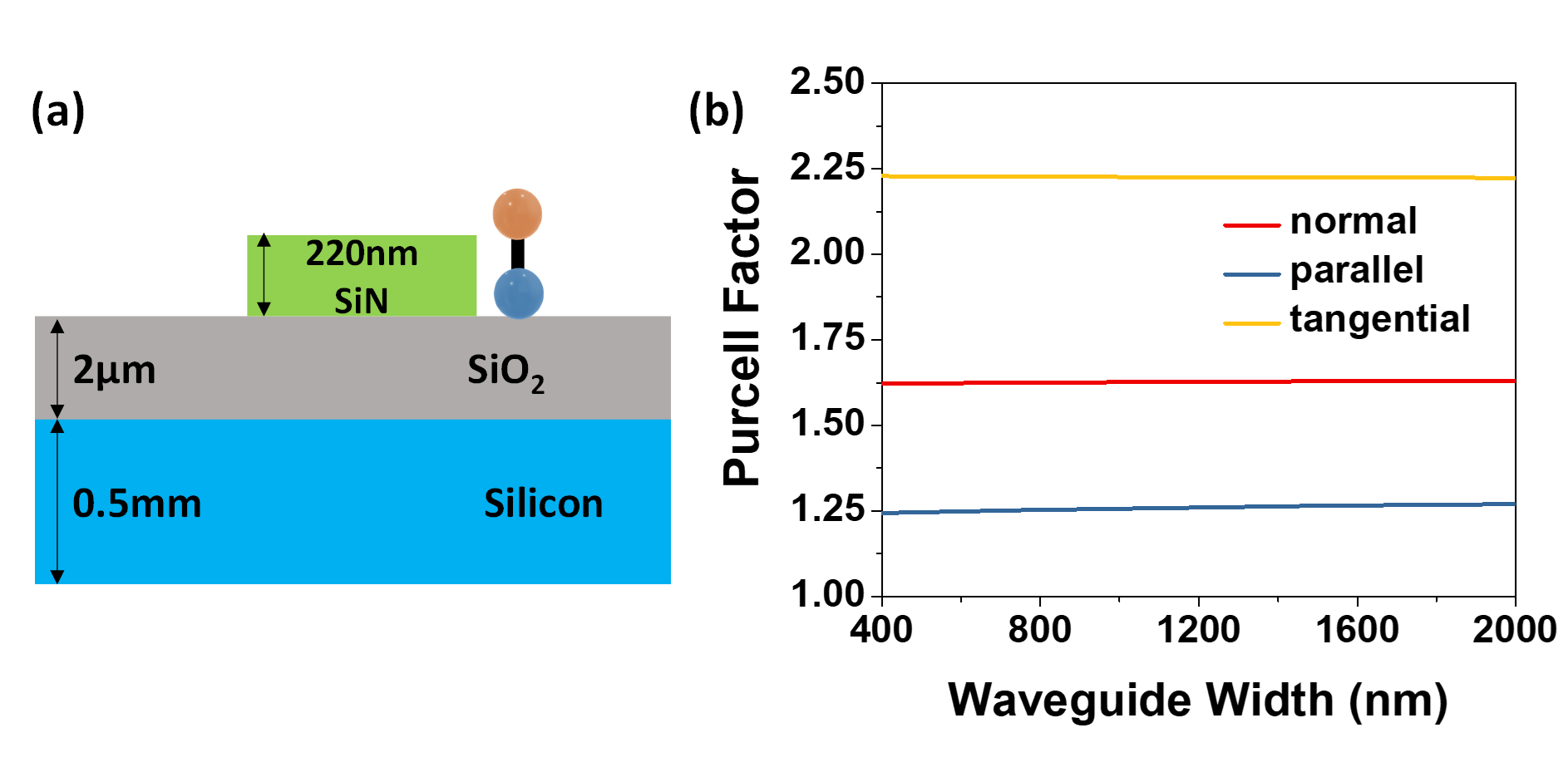}
\caption{(a) Cross section of SiN waveguide with dipole emitter at its side wall
 and (b) Purcell factors of various dipole orientations with change in waveguide width.
}
\label{fig:3}
\end{figure}
The effect of the emission wavelength of dipole emitter on the Purcell enhancement is also studied over a broad wavelength range of 500 to 1000 nm for a fixed SiN wire waveguide dimensions of 400 nm x 220 nm (figure 4(a)). Also, when the dipole emitter is far from the waveguide, i.e., away from an inhomogeneous medium, Purcell enhancement disappears after a certain distance as the dipole feels a homogeneous media. This effect is shown for various dipole orientations in Figures 4(b),4(c), and 4(d). One can see that as the proximity between the waveguide and dipole, the Purcell factor is unity, which means no spontaneous emission can occur. The Purcell factor depends on the wavelength of the emitter emission as described in equation (5), and the same can be observed from simulation results shown in Figure 4(a).
\begin{figure}
    \centering
    \includegraphics[scale =0.38]{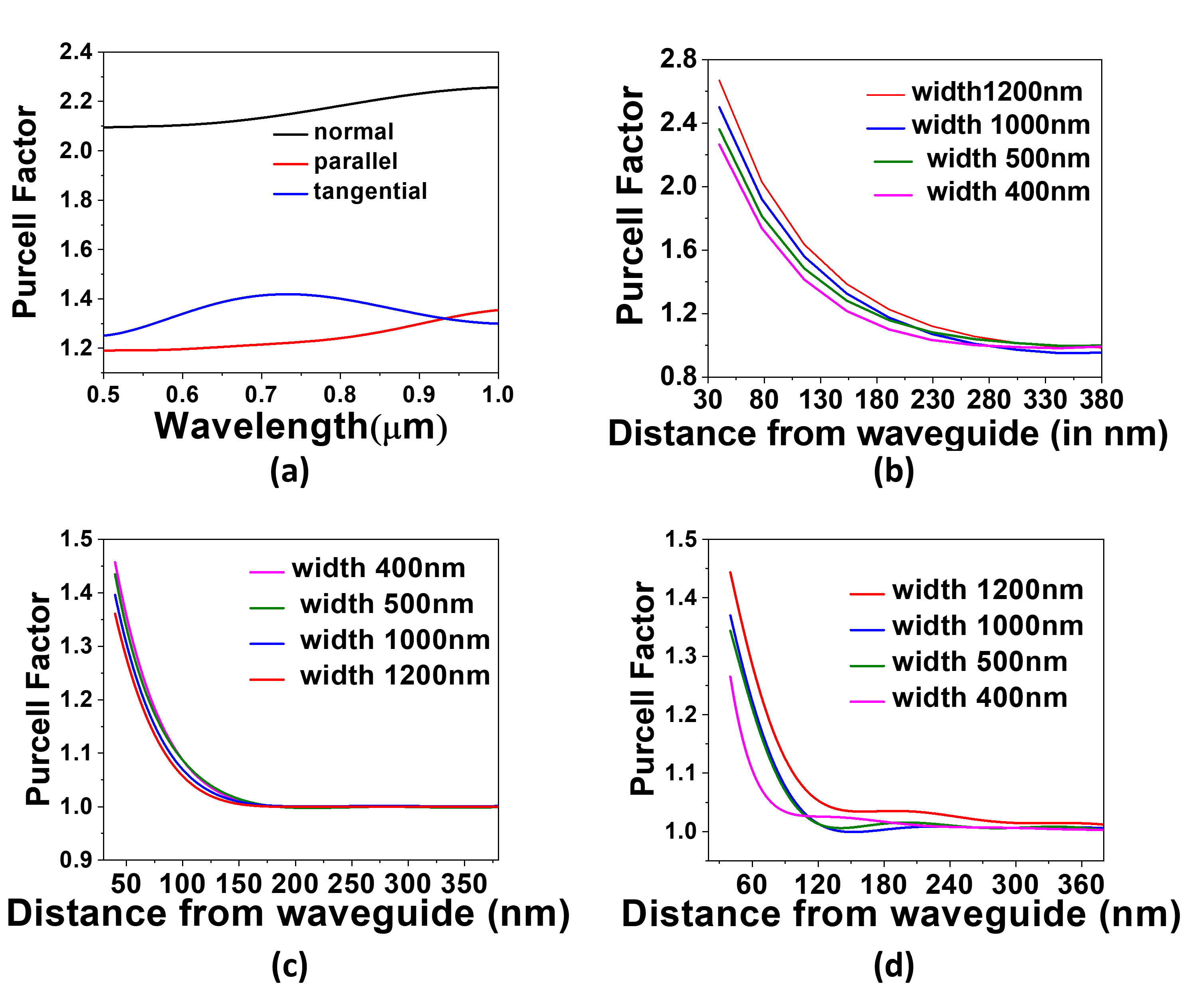}
\caption{\justifying (a) Change in Purcell factors of various dipole orientations on a 400 X 220 nm SiN waveguide as a function of wavelength. (b) Change in Purcell factor of normally oriented dipole emitter as a function of proximity to the waveguide. (c) Change in Purcell factor of parallelly oriented dipole emitter as a function of proximity to the waveguide. (d) Change in Purcell factor of tangentially oriented dipole emitter as a function of proximity to the waveguide.}
\label{fig:4}
\end{figure}

\section{Sample preparation}
In our experimental setup, we opted for a comprehensive demonstration, incorporating a fully etched wire silicon nitride (SiN) waveguide with dimensions of 400nm x 220 nm alongside a rib waveguide measuring 400nm in thickness and shallow-etched for 200nm. The substrate comprises a bare silicon substrate, meticulously cleaned, followed by depositing a 2 µm layer of silicon dioxide using Plasma-Enhanced Chemical Vapor Deposition (PECVD). Subsequently, a 220nm layer of silicon nitride, chosen for its absence of autofluorescence [1], is deposited. To configure the wire waveguides, the SiN layer undergoes precise patterning using electron beam lithography and reactive ion etching. This intricate process ensures the creation of well-defined waveguide structures for our experimental investigations.
To demonstrate the effect discussed in section 1 and 2, we integrate 2D-CdSe Nanoplatelets (NPLs) as a quantum emitter with the waveguide structures. Unlike spherical quantum dots, the dipole moments of excitons are not randomly oriented but rather exhibit preferential alignment along the long axis of the platelet due to their anisotropic geometry [14]. Indeed, in many solid-state applications, CdSe NPLs are typically used in compact thin films where the orientation of the individual nanoplatelets is random. While this random orientation allows for some exploration of the unique properties of CdSe NPLs, there is also a growing interest in controlling the direction of NPL films to maximize their potential in optoelectronic and photonic devices [15],[16].  
CdSe NPLs were synthesized using the well-known hot colloidal method [13]. In a three-neck flask, 170 mg of Cadmium myristate and 14 mL of 1-Octadecene were introduced, and the mixture was degassed under vacuum at room temperature for 60 minutes. The mixture was heated to 240°C in a nitrogen environment. A 12 mg of Se dispersed in 1 mL of ODE was quickly injected into the flask. 120 mg of Cd(Ac)2.2H2O was introduced to the flask after 60 seconds. The reaction was quenched using a water bath after 10 minutes. Oleic acid was added to the reaction mixture at 80 °C as a capping agent. The as-synthesized NPLs were dispersed in hexane at room temperature and extracted using selective precipitation. 

\section{experimental results}

\begin{figure}[ht]
    \centering
    \includegraphics[scale =0.4]{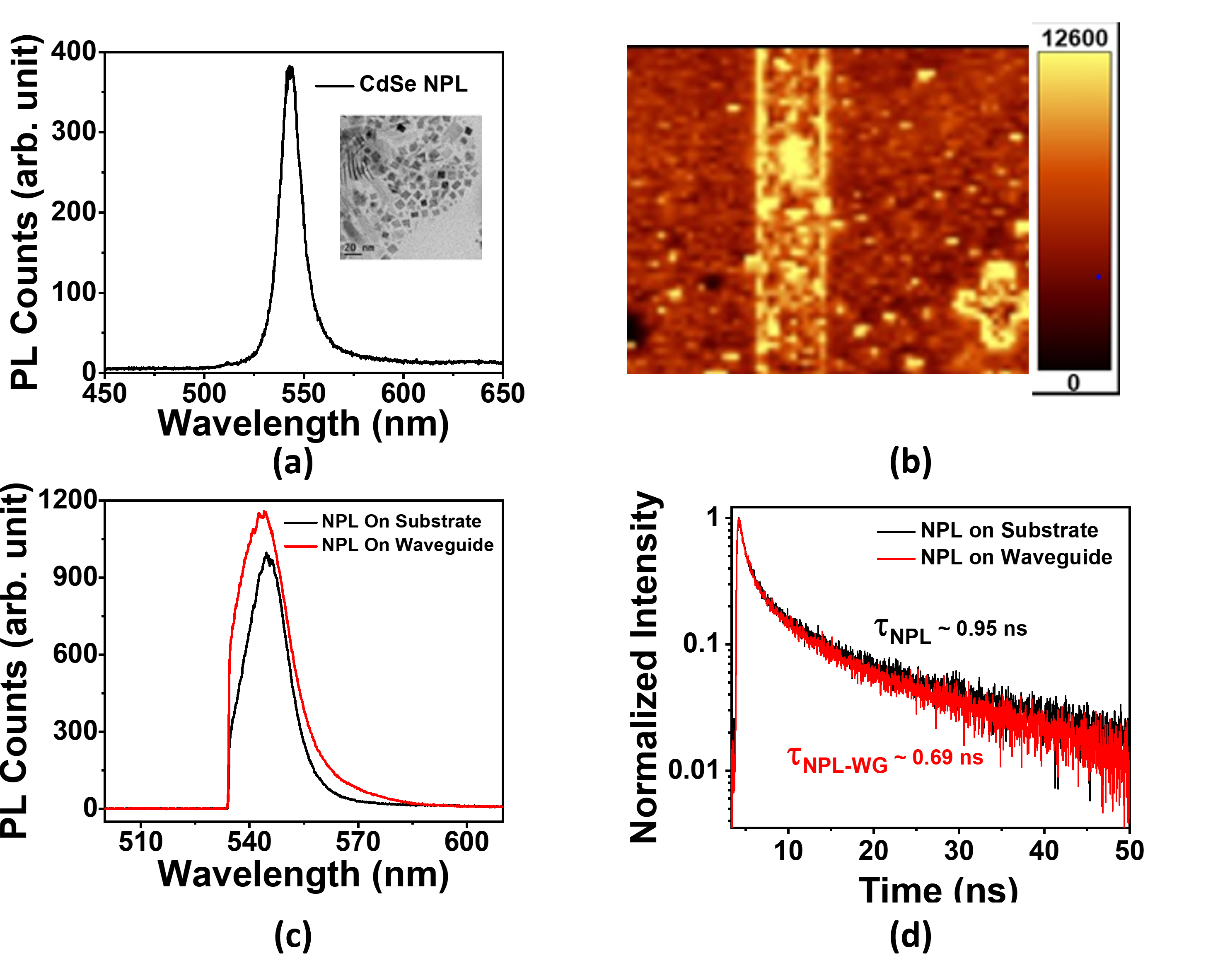}
\caption{\justifying (a) PL spectra of CdSe NPLs emitting at 543 nm. The inset shows a TEM image of the as-synthesized 2D-CdSe Nanoplatelets with a scale bar of 20 nm. (b) PL map highlighting NPL emission enhancement over the waveguide region compared to the substrate. (c) PL emission spectra of NPL on silicon (black) and waveguide (red). (d) Exciton lifetime on silicon (black) and waveguide (red).
}
\label{fig:5}
\end{figure}

Figure \ref{fig:5}a shows the Transmission Electron Microscopy (TEM) and photoluminescence (PL) characterization on the synthesized NPLs. The measurements confirm the 2D nanoparticles morphology and optical characteristics. The NPLs were dip-coated on the waveguide structure, and PL measurements were performed using laser excitation at 532 nm. The as-synthesized NPLs have an emission at 543 nm with a linewidth of 12 nm.

When placed on a waveguide structure, NPL emission gets enhanced, as observed from the PL map and intensity spectrum in Fig. \ref{fig:5}b and \ref{fig:5}c, respectively. A PL enhancement is driven by the Purcell effect, as discussed in the previous section. The enhancement factor is the ratio of PL intensity on the waveguide ($I_WG$) to on-silicon dioxide substrate($I_Si$). We observe an enhancement of about 1.22 for the NPLs on the waveguide. The enhancement agrees with the simulations for the dipole orientation aligned parallel and tangential to the waveguide, as shown in Fig. \ref{fig:4}a.  

As reported in [14], 95$\%$ of the transition dipole moment (TDM) for the CdSe NPLs is aligned in the in-plane direction (along the long axis of the NPL), suggesting the preferable orientation of the $NPL_s$ is in the face-down geometry when deposited with the dip-coating method.

To further investigate the enhancement mechanism, average exciton lifetimes on both silicondioxide and the waveguide were measured using time-resolved photoluminescence (TRPL) measurements, as presented in panel (d) of Figure 5. The exciton lifetime of NPLs placed on a silicon dioxide substrate, $\tau_{Si}$, was measured to be 0.949 ns, which decreased to 0.693 ns when placed over the waveguide structure $(\tau_{WG})$. This reduction in exciton lifetime leads to a Purcell enhancement, approximated as $\tau_{Si}⁄\tau_{WG}$, by a factor of 1.37. The simulated Purcell enhancement for those emitters aligned parallelly and tangentially to the waveguide surface is 1.225 (average of parallel and tangential orientations at $\lambda = 540 nm $ from Figure 4(a)), matching with experimental Purcell factors extracted by PL measurements, which is 1.22 and lifetime measurements which is 1.37. 
NPL emission is also studied on rib waveguides of a thickness of 400 nm and a rib thickness of 200nm. PL measurements are done on the waveguide and on the rib. As the rib can also act as a slab waveguide, the enhancement factor calculated as the ratio of on and off the waveguide is too low as shown in Figure 6(a). Exciton lifetime is also measured on and off the rib waveguide shown in Figure 6(b). The enhancement factor calculated using lifetime measurements is 1.16 and is also close to the ratio obtained by PL measurements i.e. 1.06. These findings demonstrate the potential of precisely manipulating NPL orientation and integration into photonic devices to maximize their performance and enable new functionalities.
\begin{figure}[ht]
    \centering
    \includegraphics[scale=0.42]{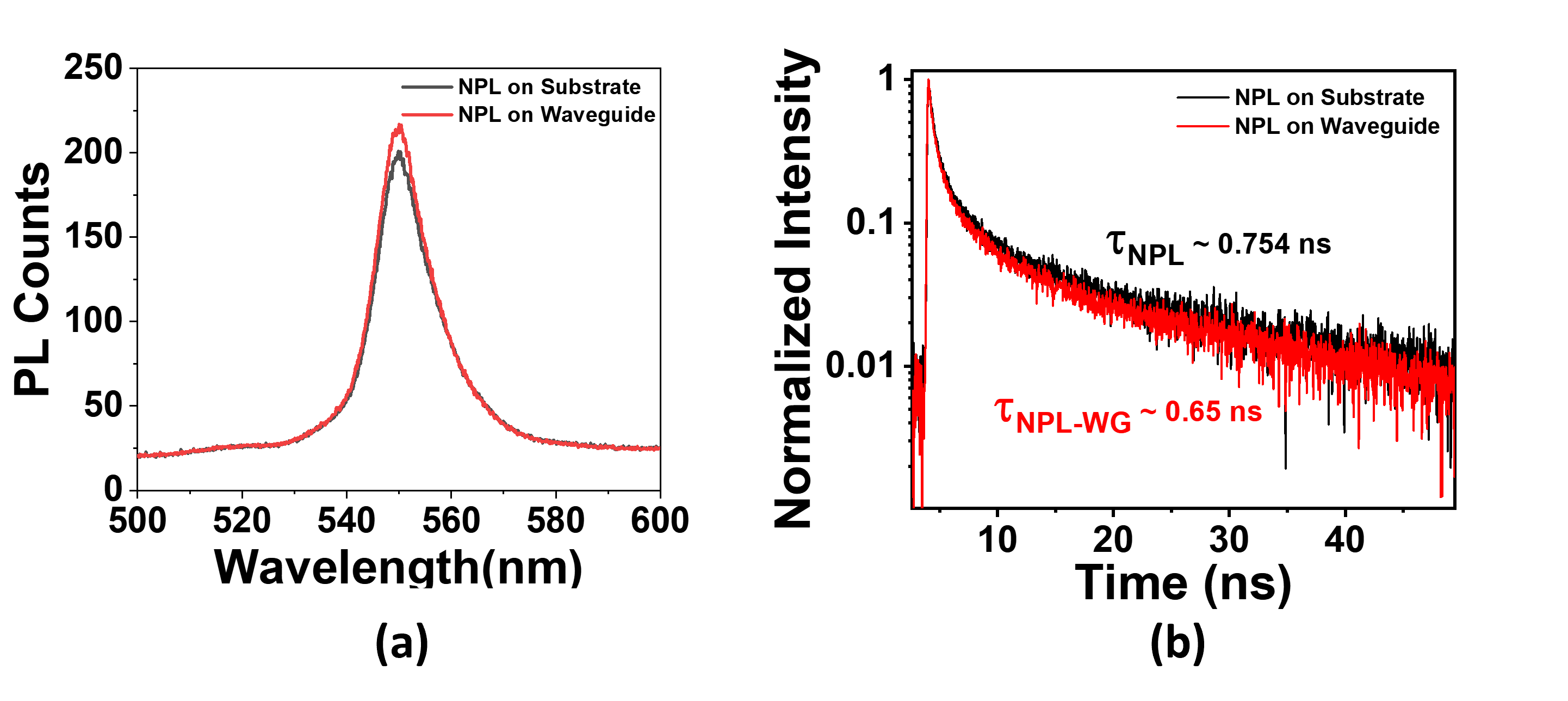}
\caption{(a) PL emission spectra of NPLs on and off rib waveguide (b)Exciton lifetime on and off waveguide
}
\label{fig:6}
\end{figure}

\section{Conclusion}
Our investigation focused on understanding the alterations in the spontaneous emission behavior of a quantum emitter when situated near a waveguide structure. Through an extensive simulation study, we explored various emitter orientations on the waveguide, assessing how both orientation and distance from the waveguide impact spontaneous emission. The simulations revealed Purcell enhancement when the emitter is in close proximity to the waveguide, and this enhancement was validated through experimental demonstrations. This study serves as a compelling proof of concept, demonstrating the Purcell enhancement offered by waveguides. The findings are instrumental in quantifying the effects of waveguides for on-chip waveguide-based spectroscopy and quantum photonic applications, providing valuable insights for practical implementations in these domains.
\begin{backmatter}

\bmsection{Acknowledgments} We acknowledge the support from MHRD, MeitY, and DST through NNetRA. We also thank UGC for the research fellowship and Prof. Rama Krishna Rao's fellowship.

\bmsection{Disclosures} The authors declare no conflicts of interest.

\bmsection{Data Availability Statement} Data underlying the results presented in this paper are not publicly available but may be obtained from the authors upon reasonable request.

\bmsection{Supplemental document}
See Supplement 1 for supporting content. 

\end{backmatter}
\noindent \cite{Zhang:14,dhakal2014evanescent,purcell1995spontaneous,Kuhn,kepler,PhysRevLett.50.1903,10.1143/ptp/5.4.570,Jun:09,1975wi...book.....K,barnes1998fluorescence,ba2012enhanced,lukas,doi:10.1021/ja110046d,doi:10.1021/acs.nanolett.8b03422,gao2017cdse,liu2020fourier}.
\bibliography{sample}

\begin{thebibliography}{10}
\newcommand{\enquote}[1]{``#1''}

\bibitem{Zhang:14}
S.~Gali, V.~Raghunathan, and S.~Kumar, \enquote{Silicon nitride waveguide platform for on-chip spectroscopy at visible and nir wavelengths,} {\protect\JournalTitle{Integrated Optics: Devices, Materials, and Technologies XXIV, SPIE}} \textbf{11283}, 158--164 (2020).

\bibitem{dhakal2014evanescent}
A.~Dhakal, A.~Z. Subramanian, P.~Wuytens, F.~Peyskens, N.~Le~Thomas, and R.~Baets, \enquote{Evanescent excitation and collection of spontaneous raman spectra using silicon nitride nanophotonic waveguides,} {\protect\JournalTitle{Optics letters}} \textbf{39}, 4025--4028 (2014).

\bibitem{purcell1995spontaneous}
E.~M. Purcell, \enquote{Spontaneous emission probabilities at radio frequencies,} {\protect\JournalTitle{Phys.Rev.Lett}} \textbf{69}, 681 (1946).

\bibitem{Kuhn}
H.~Kuhn, \enquote{Classical aspects of energy transfer in molecular systems,} {\protect\JournalTitle{The Journal of Chemical Physics}} \textbf{53}, 101--108 (2003).

\bibitem{kepler}
D.Kleppner, \enquote{Inhibited spontaneous emission,} {\protect\JournalTitle{Phys.Rev.Lett}} \textbf{47}, 233--236 (1981).

\bibitem{PhysRevLett.50.1903}
P.~Goy, J.~M. Raimond, M.~Gross, and S.~Haroche, \enquote{Observation of cavity-enhanced single-atom spontaneous emission,} {\protect\JournalTitle{Phys. Rev. Lett.}} \textbf{50}, 1903--1906 (1983).

\bibitem{10.1143/ptp/5.4.570}
E.~Fermi, \enquote{{High Energy Nuclear Events},} {\protect\JournalTitle{Progress of Theoretical Physics}} \textbf{5}, 570--583 (1950).

\bibitem{Jun:09}
Y.~C. Jun, R.~M. Briggs, H.~A. Atwater, and M.~L. Brongersma, \enquote{Broadband enhancement of light emission in silicon slot waveguides,} {\protect\JournalTitle{Opt. Express}} \textbf{17}, 7479--7490 (2009).

\bibitem{1975wi...book.....K}
J.~A. {Kong}, \emph{{Theory of electromagnetic waves}} (1975).

\bibitem{barnes1998fluorescence}
W.~Barnes, \enquote{Fluorescence near interfaces: the role of photonic mode density,} {\protect\JournalTitle{journal of modern optics}} \textbf{45}, 661--699 (1998).

\bibitem{ba2012enhanced}
T.~Ba~Hoang, J.~Beetz, L.~Midolo, M.~Skacel, M.~Lermer, M.~Kamp, S.~H{\"o}fling, L.~Balet, N.~Chauvin, and A.~Fiore, \enquote{Enhanced spontaneous emission from quantum dots in short photonic crystal waveguides,} {\protect\JournalTitle{Applied Physics Letters}} \textbf{100} (2012).

\bibitem{lukas}
B.~H. Lukas~Novotny, \emph{Principles of Nano-Optics} (Cambridge University Press, 2006).

\bibitem{doi:10.1021/ja110046d}
S.~Ithurria, G.~Bousquet, and B.~Dubertret, \enquote{Continuous transition from 3d to 1d confinement observed during the formation of cdse nanoplatelets,} {\protect\JournalTitle{Journal of the American Chemical Society}} \textbf{133}, 3070--3077 (2011). PMID: 21323349.

\bibitem{doi:10.1021/acs.nanolett.8b03422}
J.~M. Winkler, F.~T. Rabouw, A.~A. Rossinelli, S.~V. Jayanti, K.~M. McPeak, D.~K. Kim, B.~le~Feber, F.~Prins, and D.~J. Norris, \enquote{Room-temperature strong coupling of cdse nanoplatelets and plasmonic hole arrays,} {\protect\JournalTitle{Nano Letters}} \textbf{19}, 108--115 (2019). PMID: 30516054.

\bibitem{gao2017cdse}
Y.~Gao, M.~C. Weidman, and W.~A. Tisdale, \enquote{Cdse nanoplatelet films with controlled orientation of their transition dipole moment,} {\protect\JournalTitle{Nano Letters}} \textbf{17}, 3837--3843 (2017).

\bibitem{liu2020fourier}
J.~Liu, L.~Guillemeney, A.~Choux, A.~Ma{\^\i}tre, B.~Ab{\'e}cassis, and L.~Coolen, \enquote{Fourier-imaging of single self-assembled cdse nanoplatelet chains and clusters reveals out-of-plane dipole contribution,} {\protect\JournalTitle{ACS photonics}} \textbf{7}, 2825--2833 (2020).

\end{thebibliography}

\bibliographyfullrefs{sample}


\end{document}